\documentclass[twocolumn,showpacs,preprintnumbers,amsmath,amssymb,pre]{revtex4}
\usepackage{epstopdf}
\usepackage{graphicx} 
\usepackage{dcolumn} 
\usepackage{bm} 

\catcode`\^^M=10      

\begin{document}

\preprint{APS/123-QED}

\title{Rocking and rolling: a can that appears to rock might actually roll}

\author{Manoj Srinivasan$^{1}$}
 \email{msriniva@princeton.edu}
 \homepage{http://www.princeton.edu/~msriniva}
\affiliation{%
$^{1}$Mechanical and Aerospace Engineering, Princeton University, NJ 08544 \\
}%

\author{Andy Ruina$^{2}$}
 \email{ruina@cornell.edu}
 \homepage{http://ruina.tam.cornell.edu}
\affiliation{%
$^{2}$Theoretical and Applied Mechanics, Cornell University, NY 14853 \\
}%

\date{February 11, 2007}

\begin{abstract}
A beer bottle or soda can on a table, when slightly tipped and released, falls to an upright position and then rocks up to a somewhat opposite tilt.  Superficially this rocking motion involves a collision when the flat circular base of the container slaps the table before rocking up to the opposite tilt.  A keen eye notices that the after-slap rising tilt is not
generally just diametrically opposite the initial tilt but is veered to one side or the other.   Cushman and Duistermaat (2006) recently noticed such veering
when a flat disk with rolling boundary conditions is dropped nearly flat.  Here,
we generalize these rolling disk results to arbitrary axi-symmetric bodies and to frictionless sliding.  More specifically,  we study motions that almost but do not quite involve a face-down collision of the round container's bottom with the table-top. These motions involve a sudden rapid motion of the contact point around the circular base. Surprisingly, like for the
rolling disk,  the net angle of motion of this contact point is nearly independent of initial conditions.  This angle of turn depends simply on the geometry and mass distribution but not on the moment of inertia about the symmetry axis. We derive simple asymptotic formulas for this ``angle of turn'' of the contact point and check the result with numerics and with simple experiments. For tall containers (height much bigger than radius) the angle of turn is just over $\pi$ and the sudden rolling motion superficially appears as a nearly symmetric collision leading to  leaning on an almost diametrically opposite point on the bottom rim.
\end{abstract}

\maketitle

\section{Introduction}
\begin{figure}[t]
	\centering
		  \includegraphics[width=\hsize]{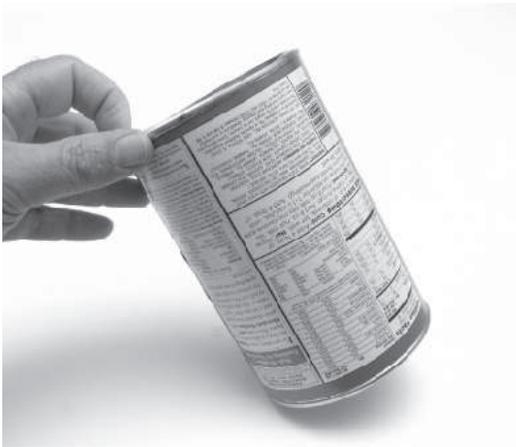}
	\caption{\textbf{Releasing a tipped container.} After this release the
	container falls and rocks as shown in Figs.~\ref{fig:fallsequence} and
	\ref{fig:scatterplotovaltine}.}
	\label{fig:OvaltineTippedPhoto}
\end{figure}

\begin{figure}
	\centering
		\includegraphics{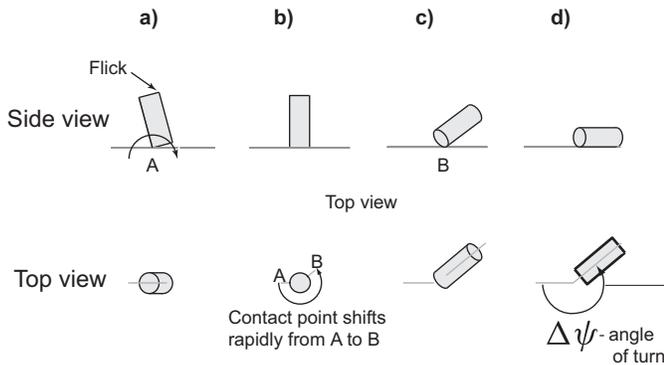}
	\caption{\textbf{Toppling a container.} Schematic time-sequence of container (illustrated
 as a cylinder) during one tipping experiment: side view (top row) and top view (2nd row).  a) Tilt a container  and flick its top with your finger, so that b) the container becomes vertical again, and then c), d) tips part way over some more
	and then eventually  d)  has sufficient energy to fall on its  side. The container hardly ever falls symmetrically to the diametrically opposite side with $\Delta\psi=\pi$. 
	See Fig.~\ref{fig:scatterplotovaltine}}
	\label{fig:fallsequence}
\end{figure}

After a meal or a drink, fidgety people sometimes play with the available props, such as empty glasses, bottles, and soda cans. These containers are all nearly axi-symmetric objects with round bottoms. A natural form of play is to roll these containers on their
circular bottoms.  When a slightly tipped container (Fig.~\ref{fig:OvaltineTippedPhoto}) is  let go, sometimes one sees  it fall upright, make a slight banging sound, and then tip up again at another angle, then fall back upright again, and so on for a few rocking oscillations. Because these oscillations usually damp quickly its  easy to miss the details of the motions.  To aid the eye, instead  of just letting go of the slightly tilted container, you can flick its top forward with the fingers. This provides an initial righting angular velocity  along the axis about which the container was initially tipped  (too large an angular velocity will cause the container to lift off the table as it pivots over). The tipping container, as before, falls to a vertical configuration, at which time  its bottom circular face bangs on the table. After this bang, the container tips up onto the other side and maybe falls over. When this experiment is performed with a container that is not too tall and too thin, you can see that  the container does not fall exactly onto the diametrically opposite side of the bottom rim. That is, point A on the bottom of the container that initially contacted the
table  and the new contact point B, that the container rocks up onto, are not exactly 180 degrees apart.  This experiment (video available online \footnote{A video of this experiment is available from the authors' web pages and as a supplementary EPAPS file from the APS website.}) is shown schematically in Fig.~\ref{fig:fallsequence}. Fig.~\ref{fig:scatterplotovaltine} shows a histogram of one particular container's orientations after we repeatedly flicked it. The distribution is strongly bimodal with no symmetric falls over many repeated trials.  

\par
Note the apparent symmetry breaking with $\Delta\psi\ne\pi$. Is this deviation from symmetric rocking due to imperfect hand release? Here we show that the breaking of apparent symmetry is consistent with the simplest deterministic theories, namely smooth rigid body dynamics. We derive formulas for the ``angle of turn'', for both the perfect-rolling and for the frictionless-sliding cases given infinitesimal symmetry breaking in the initial conditions.  The results here generalize some of the nearly-falling-flat results of Cushman and Duistermaat \cite{Cus06}; they considered the special case of the pure-rolling of flat disks.

\section{Candidate theories for symmetry breaking}
If the table was perfectly planar and horizontal, the container's bottom was perfectly circular, and the container was perfectly axisymmetric, then an initial condition with a purely righting angular velocity would result in a collision in which, just as the container becomes vertical, all points on the container's bottom slap the table-top simultaneously. The consequence of such a rigid-container collision is not computable since algebraic rigid-body collision laws are not well-defined for simultaneous multi-point collisions; for example, the order of the
impulse locations is then ill-defined \cite{Ivan95,Goy98,Chat98,Rui05}.  Basing the rocking outcome on such a perfect flat collision would be basing the outcome on details of the deformation, that is, leaving the world of rigid-object mechanics.

\par
\begin{figure}
	\centering
	   	  		\includegraphics{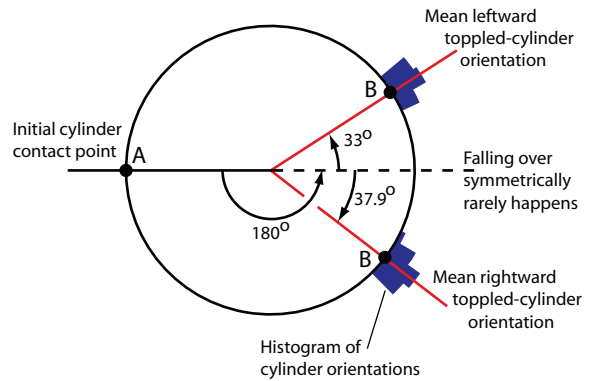}
	\caption{\textbf{Typical toppled container orientations.} A histogram of the directions along which a container fell for 42 trials. The container always falls to one side or another, but essentially never to the diametrically opposite side corresponding to 180 degrees.  The distribution about $\Delta\psi =180$ degrees is strongly bimodal. Whether the fall is to the left or to the right depends sensitively on initial conditions. The leftward falls have a mean of about $\Delta\psi =180+33$ degrees and the rightward falls have a mean of about $\Delta\psi =180-37.9$ degrees. For this cylinder theory predicts $\Delta\psi =180\pm 40$ degrees for frictionless sliding and $\Delta\psi =180\pm 22$ degrees for rolling with no slip.}
	\label{fig:scatterplotovaltine}
\end{figure}

 Geometric perturbations to the circular rim can similarly lead to two-point collisions, where both points of contact are now on the rim. The second point of contact could be diametrically opposite to the first point, or anywhere else on the rim. We could then compute the consequences of, say, a plastic collision at the new point of contact. However, the consequences of the collision would depend on the location of the geometric imperfection. For a given collision point, such a theory could predict the energy dissipation at the collision. But such a theory cannot be useful in predicting a systematic breaking of symmetry as seen in Fig.~\ref{fig:scatterplotovaltine}. For any given imperfection, the motion is deterministic and depends on the location of the impection, and there is no reason to expect that the location of the imperfection would have a distribution similar to that in Fig.~\ref{fig:scatterplotovaltine}.

\par
Assuming a perfectly flat rigid bottom, the circle-slapping-ground simultaneous collision is essentially impossible. After all, the container is being launched by imperfect human hands that cannot provide any
exact initial conditions.  Accounting for various symmetries in the problem, the set of all motions of a container rolling without slip is three dimensional; the space of solutions could be parameterized by, say, the minimum tip angle $\phi_{min}$, the yaw rate $\dot{\psi}$ at that position, and the rolling rate $\dot{\theta}$ at that position. The set of solutions that leads to a face-down collision can be characterized with only two parameters, a set with co-dimension one \cite{Orei96,Bor03,Cus06}. So, small generic perturbations of a ``collisional'' initial condition result, with probability 1,  in a non-collisional motion described by  the smooth dynamics of the container rolling or sliding on its circular bottom rim.

\par
The rest of this paper is about the near-collisional motions.
We will see that these near-collisional motions involve rapid  rolling or sliding of the container on its bottom rim, so that the contact point appears to have switched by a finite angle that is greater than 180 degrees.  For this analysis, we assume a geometrically  perfect container (axisymmetric) and table (flat).

\section{Dynamics of Rocking by rolling}
Consider a container with mass $m$, bottom radius $R$ \footnote{Only the bottom radius is important. The container need not be a constant-radius cylinder.}, and the center of mass at a height $H$ from the bottom. The moment of inertia is $C$ about it's symmetry axis and is $A$ about any axis passing through the center of mass and perpendicular to the symmetry axis. For the disk of \cite{Cus06}, $H=0$ and $A=C/2$.  In our case, $H\ge0$ and $A\ge C/2$.

\par
\begin{figure}
	\centering
		\includegraphics[scale=1.1]{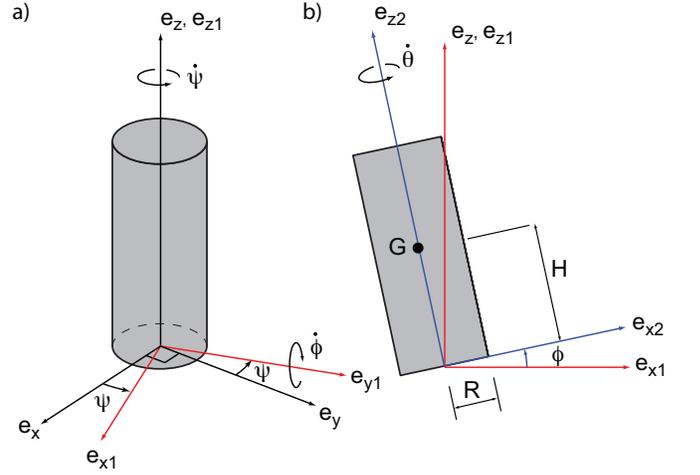}
	\caption{Definition of the Euler angles and coordinate axes used to define the orientation of the container.}
	\label{fig:eulerangles}
\end{figure}

The center of mass of the container is at $(x_G,y_G,z_G)$ in an inertial frame $\mathbf{e}_x$-$\mathbf{e}_y$-$\mathbf{e}_z$ at rest with respect to the table. The reference orientation of the container is vertical as shown in Fig.~\ref{fig:eulerangles}a. Any other orientation of the container can be obtained from the reference orientation by a sequence of three rotations, defining corresponding Euler angles as illustrated in Fig.~\ref{fig:eulerangles}. The first rotation (`yaw' or `steer') is about $\mathbf{e}_z$ by an angle $\psi$. This rotation also transforms the inertial coordinate axes to $\mathbf{e}_{x1}$-$\mathbf{e}_{y1}$-$\mathbf{e}_{z1}$. The second rotation (`pitch' or `tilt') by an angle $-\phi$ about the $\mathbf{e}_{y1}$ axis results in the $\mathbf{e}_{x2}$-$\mathbf{e}_{y2}$-$\mathbf{e}_{z2}$ frame and determines the orientation of the container up to a rotation $\theta$ about the body-fixed symmetry axis ($\mathbf{e}_{z2}$ axis). The angular velocity of the container in the rotating $\mathbf{e}_{x2}$-$\mathbf{e}_{y2}$-$\mathbf{e}_{z2}$ frame is therefore entirely along the symmetry axis $\mathbf{e}_{z2}$ and this relative angular velocity magnitude is denoted by $\dot{\theta}$.


\par
We will mostly consider two simple extremes for the frictional interaction between the table and the container's bottom, namely, sliding without friction and rolling without slip. 

\par
For pure rolling, the Euler angles and their first
and second time derivatives determine the center-of-mass position (relative to the
contact point), the center-of-mass  velocity and acceleration,
as well as the container's angular velocity and angular acceleration.   The three second
order ODEs that determine the evolution of the orientation $\psi$, $\phi$, and $\theta$ follow
from angular momentum balance about the contact point:

\begin{equation}
Q_{i1}\ddot{\psi}+Q_{i2}\ddot{\phi}+Q_{i3}\ddot{\theta} = S_i, \quad i = 1,2,3. \label{eq:generalcyl}
\end{equation}

where

\begin{eqnarray}
Q_{11} &=& A\sin{\phi}-mHR\cos{\phi}+mH^2\sin{\phi}, \nonumber \\
Q_{12} &=&  0, \ Q_{13} = -mHR,\nonumber \\ 
Q_{21} &=& Q_{23} = 0, \ Q_{22} = -mR^2-mH^2-A, \nonumber \\
Q_{31} &=& C\cos{\phi} + mR^2\cos{\phi}-mRH\sin{\phi}, \nonumber \\
Q_{32} &=& 0, \ Q_{33} = C+mR^2, \nonumber \\
S_1 &=& (C-2A-2mH^2)\dot{\psi}\dot{\phi}\cos{\phi} +C\dot{\phi}\dot{\theta} \label{eq:noslip_eqs} \\
    &-&2mHR\dot{\psi}\dot{\phi}\sin{\phi} \nonumber \\
S_2 &=& (C-A+mR^2-mH^2)\dot{\psi}^2\sin{2\phi}/2 \nonumber \\ 
   &+& (C+mR^2)\dot{\theta}\dot{\psi}\sin{\phi} + mHR\dot{\psi}\dot{\theta}\cos{\phi} \nonumber \\ 
 &+& mHR\dot{\psi}^2\cos{2\phi} \nonumber \\
 &+& mg(R\cos{\phi}-H\sin{\phi}), \nonumber \\
S_3 &=& C\dot{\psi}\dot{\phi}\sin{\phi} + 2mR\dot{\psi}\dot{\phi}(R\sin{\phi}+H\cos{\phi}). \nonumber
\end{eqnarray}

These equations are derived in Appendix~\ref{app:eqmotion}.
Because the center-of-mass velocity  is  determined by the Euler angles and their rates, once
these are known the center of mass position can be found by integration.

\par
For the frictionless sliding of the cylinder on its circular bottom, the equations of motion are also given by Eq.~\ref{eq:generalcyl} but with
\begin{eqnarray}
Q_{11} &=& A\sin{\phi}, \ Q_{12} = Q_{13} = Q_{21} = 0, \nonumber \\
Q_{22} &=& -A+mHR\sin{2\phi}-mH^2\sin^2{\phi}-mR^2\cos^2{\phi}, \nonumber \\
Q_{23} &=& 0, \ Q_{31} = C\cos{\phi}, \ Q_{32} = 0, \ Q_{33} = C, \nonumber \\
S_1 &=& -2A\dot{\psi}\dot{\phi}\cos{\phi}+C\dot{\phi}\dot{\theta}+C\dot{\phi}\dot{\psi}cos{\phi}, \label{eq:nofric_eqs} \\
S_2 &=& (1/2)(C-A)\dot{\psi}^2\sin{2\phi} \nonumber \\ 
   &+& C\dot{\theta}\dot{\psi}\sin{\phi}+ mgR\cos{\phi}-mgH\sin{\phi} \nonumber \\ 
   &-& mHR\dot{\phi}^2\cos{2\phi}+(1/2)m(H^2-R^2)\dot{\phi}^2\sin{2\phi}, \nonumber \\
S_3 &=& C\dot{\psi}\dot{\phi}\sin{\phi}. \nonumber
\end{eqnarray}

\par
When the table is frictionless, the horizontal velocity of the center of mass is a constant, so that the horizontal position is independent of the orientation. The vertical position of the center of mass is simply given by $z_G = H\cos{\phi}+R\sin{\phi}$.  Without loss of generality
the center-of-mass can be taken as on a fixed vertical line.

\par
Our initial discovery of the phenomenon described here was found in numerical simulations
of the equations above.  We used an adaptive time-step, stiff integrator because
the solutions of interest have vastly changing time-scales. The singularity of the Euler-angle description in the near-vertical configuration also contributes to the stiffness of the equations.

\begin{figure}
	\centering
		\includegraphics{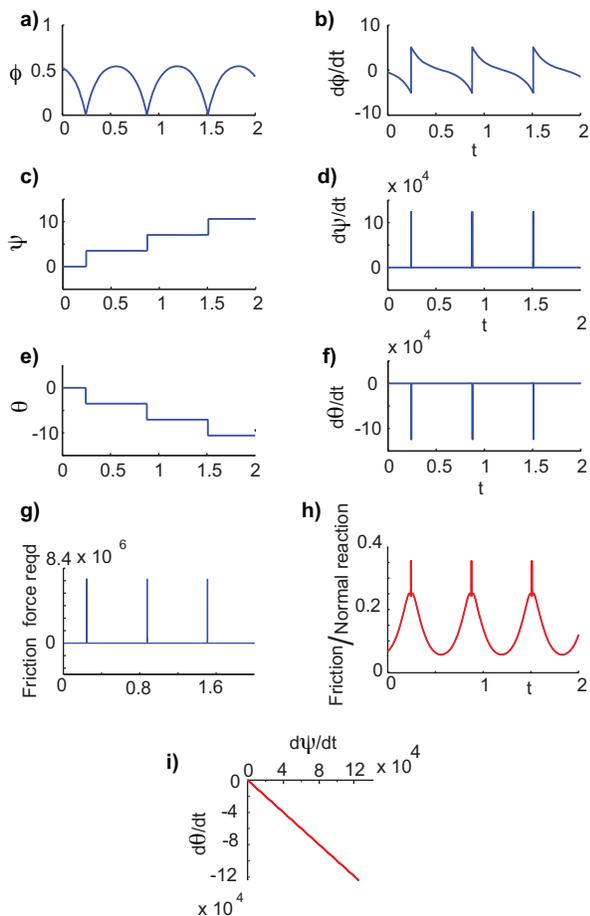}
	\caption{Results of a typical simulation of a container's face almost but not quite falling flat on the table-top. Physical parameters correspond to the can shown in Fig.~\ref{fig:OvaltineTippedPhoto} and used for the experiments in Fig.~\ref{fig:scatterplotovaltine}: $m=0.070; A=5.1 \times 10^{-3}; C=8 \times 10^{-3}; g=9.8; R = 5.1 \times 10^{-2}; H = 6.9 \times 10^{-2}$ in consistent SI units. Initial conditions $\psi(0) = 0, \dot{\psi}(0) =0 .005,  \phi(0) = \pi/6,  \dot{\phi}(0) = -0.5,  \theta(0)=0, \dot{\theta}(0) = 0.005$. All the angles are in radians. The step-change in $\psi$ is the angle of turn $\Delta \psi$, and is a little more than $\pi$ as noted in the caption for Fig.~\ref{fig:scatterplotovaltine}.}
	\label{fig:purerollnumresults}
\end{figure}

\par
For simplicity of presentation, we first describe in detail the results of integration when the container does not slip with respect to the table. First, we integrate the equations with initial conditions that lead exactly to a face-down collision $\dot{\psi}(0)= \dot{\theta}(0) = 0$, $\dot{\phi} < 0$, and say, $0 < \phi < \pi/2$. As would be expected, the container lands in a manner that all of its bottom face simultaneously reaches the horizontal table-top. Up to this time the motion of the container is identical to that of an inverted pendulum hinged at the contact point A.  The integration becomes physically meaningless at that collision point.

\par
In the vicinity of these collisional motions, the solutions of the equations of motion here do not have smooth dependence of solutions on initial conditions. To obtain a near-collisional motion, we set $\dot{\psi}(0)= O(\epsilon)$, $\dot{\theta}(0) = O(\epsilon)$, $\dot{\phi} < 0$, and $0 < \phi(0) < \pi/2$, where $\epsilon$ is a small quantity. The results of the integration are shown in Fig.~\ref{fig:purerollnumresults}. The plot of $\phi(t)$ suggests a motion in which the container's bottom face periodically comes close to touching the table ($\phi \rightarrow 0$), but then gets ``repelled'' by the floor as if by an elastic collision, so that the container rocks down and up periodically, ad infinitum, without losing any energy, as  expected from this dissipation-free system. We notice that when $\phi \approx 0$, $\dot{\phi}$ changes almost discontinuously. Also, when $\phi$ is close to zero, both $\dot{\psi}$ and $\dot{\theta}$ blow up to very large values, resulting in almost discontinuous changes in the corresponding angles $\psi$ and $\theta$. 

\par
Note that we are simply simulating the apparently smooth differential equations, and not applying an algebraic transition rule for a collision. As $\epsilon$ goes to zero, the angle rates $\dot{\psi}$ and $\dot{\theta}$ grow without bound, but the magnitude of the angular velocity vector is always bounded, as it must be since energy remains a constant throughout the motion. In particular, while the angle rates $\dot{\theta}$ and $\dot{\psi}$ are large, they are very close to being equal and opposite ($\dot{\theta} \approx -\dot{\psi}$, Fig.~\ref{fig:purerollnumresults}i). 

\par
Let us examine the consequences of a rapid finite change in $\psi$ when $\phi \approx 0$. The position of the contact point P$(x_P,y_P,z_P)$ on the ground relative to the center of mass $(x_G,y_G,z_G)$ is given by the following equations:
\begin{eqnarray}
x_P &=& x_G - R\cos{\phi}\cos{\psi}+H\sin{\phi}, \nonumber \\
y_P &=& y_G - R\sin{\psi}, \nonumber \\
z_P &=& z_G - R\sin{\phi}\cos{\psi} - H\cos{\phi} = 0. \nonumber
\end{eqnarray}
When $\phi \approx 0$ and $\psi$ is not particularly close to any multiple of $\pi/2$, the contact point position is given by 
\begin{eqnarray}
x_P &\approx& x_G - R\cos{\psi} \nonumber \\
y_P &\approx& y_G - R\sin{\psi} \nonumber \\
z_P &\approx& z_G - H = 0 \label{eq:contactptcircle}
\end{eqnarray}

\par
Given that $x_G, y_G, z_G$ do not change much during the brief near-collisional phase (because center of mass velocity is finite), we can see from Eq.~\ref{eq:contactptcircle} that a rapid continuous change in $\psi$ corresponds to a rapid continuous change in the contact point in a circle with the center $O(x_G,y_G,0)$. Thus the ``angle of turn'' AOB defined earlier is simply the change in $\psi$. At the singular limit of a near-collisional motion arbitrarily close to a collisional motion, this continuous but steep change in $\psi$ approaches a step change -- this is the ``limiting angle of turn'' and we denote this by $\Delta \psi_{turn}$.

\par
We consider the near-collisional motions of the no-slip container that can be characterized as being the pasting-together of two qualitatively distinct motions of vastly different time-scales:
\begin{enumerate}
\item inverted pendulum-like motion about an essentially fixed contact point when the tilt angle $\phi$ is large.
\item rapid rolling of the container which accomplishes in infinitesimal time, a finite change in the contact point, and a sign-change in the tilt rate $\dot{\phi}$.
\end{enumerate}

\par
Note that the near-collisional motion for a container rolling without slip requires very high friction forces (for non-zero $H$, see Fig.~\ref{fig:purerollnumresults}g). However, plotting the ratio of the required friction forces with the normal reaction, we find that only a finite coefficient of friction is required for preventing slip even in the collisional limit (Fig.~\ref{fig:purerollnumresults}h).

\par
The other extreme of exactly zero friction is similar. Here, the horizontal component of the center of mass velocity may be taken to be zero. The near collisional motions for a container sliding without friction are again characterized as consisting of two qualitatively different phases:
\begin{enumerate}
\item a tipping phase when $\phi$ is not too small, involving the container moving nearly in a vertical plane, the center of mass moving only vertically, and the contact point slipping without friction.
\item a rapid sliding phase in which the sign of $\dot{\phi}$ is reversed almost discontinuously, and the contact point moves by a finite angle in infinitesimal time.
\end{enumerate}

\par
The angle of turn for the frictionless case and the no-slip case are different in general (when $H \ne 0$). In the next two sections, we derive the formulas for the angle of turn by taking into account the two-phase structure of the near-collisional motion.

\section{Angle of turn for no-slip rolling}
\label{sec:noslipasymptotics}
The rigid body dynamics of disks, containers, and similar objects with special symmetries, have been discussed at length by a number of authors, including distinguished mechanicians such as Chaplygin, Appell, and Korteweg. Their works include complete analytical characterizations of the solutions to the relevant equations of motion, typically involving non-elementary functions such as the hyper-geometric. Reasonable reviews of such literature can be found, for instance, in \cite{Orei96} and \cite{Bor03}. We will not use these somewhat cumbersome general solutions but will analyze only the special near-collisional motion of interest to us.

\par
The calculation below may be called, variously, a boundary layer calculation, a matched asymptotics calculation (but we are not interested in an explicit matching) or a singular perturbation calculation. Essentially, we take advantage of the presence of two dynamical regimes with vastly different time-scales, each regime simple to analyze by itself. The overall motion can be obtained approximately by suitably pasting together the small-$\phi$ and the large-$\phi$ solutions. But the angle of turn is entirely determined by the small-$\phi$ regime, as will be seen below.

\par
First, consider Eq.~\ref{eq:generalcyl} with $i=1$ in the limit of small $\phi$, so that we can use $\sin{\phi} = \phi$ and $\cos{\phi} = 1$, and generally neglect terms of $O(\phi)$. We obtain
\begin{multline}
(A\phi -mHR+mH^2 \phi)\ddot{\psi}-mHR\ddot{\theta} =  \\
(C-2A-2mH^2)\dot{\psi}\dot{\phi}+ C\dot{\theta}\phi \label{eq:eq1atsmallphi}
\end{multline}

\par
Now, considering Eq.~\ref{eq:generalcyl} with $i = 3$ in the limit of small $\phi$, we obtain
\begin{multline}
(C+mR^2-mRH\phi) \ddot{\psi} + (C+mR^2) \ddot{\theta} = 2mRH\dot{\psi}\dot{\phi}. \label{eq:eq3atsmallphi}
\end{multline}

\par
Eqs.~\ref{eq:eq1atsmallphi} and \ref{eq:eq3atsmallphi} are linear and homogeneous in $\dot{\psi}$, $\dot{\theta}$, and their time derivatives. So, positing a linear relation between $\dot{\psi}$ and $\dot{\theta}$, we find that the following two equations are equivalent to Eqs.~\ref{eq:eq1atsmallphi} and \ref{eq:eq3atsmallphi}:
\begin{equation}
\dot{\psi}+\dot{\theta} = 0 \label{eq:psieqtheta}
\end{equation}
and 
\begin{equation}
\frac{\ddot{\psi}}{\dot{\psi}} = -2 \frac{\dot{\phi}}{\phi}. \label{eq:noslip1}
\end{equation}

Note that Eq.~\ref{eq:psieqtheta} agrees with the results of the numerical simulation, as in Fig.~\ref{fig:purerollnumresults}i. Even though this equation was derived for small $\phi$, this equation is approximately true at large $\phi$ as well if the initial conditions of the motion at large $\phi$ have $\dot{\psi}(0) = O(\epsilon)$ and $\dot{\theta}(0) = O(\epsilon)$, as is the case for the near-collisional motions we consider.

\par
Integrating Eq.~\ref{eq:noslip1}, we obtain 
\begin{equation}
\dot{\psi} = \frac{b_2}{\phi^2}. \label{eq:psidotandphi}
\end{equation}
Thus, when $\phi \ll 1$, both $\dot{\psi}$ and $\dot{\theta}$ ($= -\dot{\psi}$) become very large.

\par
Now consider Eq.~\ref{eq:generalcyl} with $i = 2$ and with $\phi \ll 1$.
\begin{multline}
-(A+mR^2+mH^2)\ddot{\phi} = \dot{\psi}^2\phi(C-A+mR^2-mH^2) \\
+ \dot{\psi}\dot{\theta}\phi(C+mR^2) + mHR\dot{\psi}\dot{\theta} + mHR\dot{\psi}^2+ mgR \label{eq:eq2atsmallphi}
\end{multline}
Using Eq.~\ref{eq:psieqtheta} in Eq.~\ref{eq:eq2atsmallphi} and ignoring higher order terms in $\phi$, we obtain
\begin{equation}
(A+mR^2+mH^2) \ddot{\phi} = (A+mH^2)\phi\dot{\psi}^2 -mgR \label{eq:ddphianddpsi}
\end{equation}

Using Eq.~\ref{eq:psidotandphi} for $\dot{\psi}$ in the above equation and neglecting the $mgR$ term (as it is a constant and therefore, much smaller than 1/$\phi^3$ for small $\phi$), we get
\begin{eqnarray}
 \ddot{\phi}	&=& \frac{A+mH^2}{A+mR^2+mH^2} \frac{b_2^2}{\phi^3} \nonumber \\
              &=& \frac{b_3}{\phi^3} \label{eq:rollphidotdoteqn},
\end{eqnarray}
where 
\begin{equation}
b_3 = \frac{A+mH^2}{A+mR^2+mH^2} b_2^2.
\end{equation}

The general solution for the differential equation in $\phi$, Eq.~\ref{eq:rollphidotdoteqn}, can be written as
\begin{equation}
\phi^2 = \phi^2_c + b_3 (t-t_c)^2/\phi^2_c \label{eq:phiandt}
\end{equation}
where $\phi_c$ is the lowest $\phi$ attained by the container before rising back up again, and $t = t_c$ is when this minimum $\phi$ is attained. Substituting this equation in Eq.~\ref{eq:psidotandphi}, we obtain a simple equation for the evolution of $\psi$ when near the surface:
\begin{equation}
\dot{\psi} = \frac{b_2}{\phi^2} = \frac{b_2}{\phi^2_c + b_3 (t-t_c)^2/\phi^2_c} \nonumber
\end{equation}
We can now compute the change in $\psi$ during a small time interval $\Delta t$ centered at $t=t_c$.
\begin{eqnarray}
\Delta \psi_{turn} &=& \int_{t_c+\Delta t}^{t_c-\Delta t} \dot{\psi} dt \nonumber \\ 
&=& \int_{t_c+\Delta t}^{t_c-\Delta t} \frac{b_2}{\phi_c^2+b_3(t-t_c)^2/\phi^2_c} dt \nonumber \\
            &=& 2\frac{b_2}{\sqrt{b_3}}\tan^{-1}\left( \frac{\sqrt{b_3}\Delta t}{\phi_c^2} \right) \nonumber \\
            &=& 2\sqrt{\frac{A+mR^2+mH^2}{A+mH^2}}\tan^{-1}\left( \frac{\sqrt{b_3}\Delta t}{\phi_c^2} \right) \nonumber 
\end{eqnarray}
From conservation of energy, we can show that $b_2$, and therefore $\sqrt{b_3}$, must be $O(\phi_c)$ (see Appendix \ref{app:b2scaling}). Hence, $\sqrt{b_3}/\phi_c^2 \rightarrow \infty$ as $\phi_c \rightarrow 0$. Keeping $\Delta t$ a small constant as we let $\phi_c \rightarrow 0$, thus approaching the collisional limit, gives us the following expression for $\Delta \psi$.
\begin{eqnarray}
\Delta \psi_{turn} &=& 2\sqrt{\frac{A+mH^2}{A+mR^2+mH^2}} \ \tan^{-1}\left( \infty \right) \nonumber \\
           &=& \pi \sqrt{\frac{A+mH^2}{A+mR^2+mH^2}} \label{eq:noslipangleofturn}
\end{eqnarray}
This is the limiting angle of turn when the container rocks by rolling without slip. Earlier numerical integrations agree quite well with this formula near the collisional limit, as they should.

\section{Angle of turn for frictionless sliding}
\label{sec:nofricasymptotics}
We now briefly outline the procedure for deriving the angle of turn for the frictionless case. The procedure closely parallels that described for the no-slip case, except for small differences below. The frictionless equations Eq.~\ref{eq:generalcyl} and Eq.~\ref{eq:nofric_eqs} corresponding to $i = 3$ simplifies to 
\begin{eqnarray}
C\frac{d}{dt}(\dot{\psi}\cos{\phi}+\dot{\theta}) &=& 0 \nonumber \\
\dot{\theta} + \dot{\psi}\cos{\phi} = \mathrm{constant } &=& b_1, \mathrm{ say.} \label{eq:slipangmom}
\end{eqnarray}
Here, $b_1 = O(\epsilon) \ll 1$ because the initial conditions satisfy $\dot{\theta}(0) = O(\epsilon)$ and $\dot{\psi}(0) = O(\epsilon)$ by assumption, as before. Using Eq.~\ref{eq:slipangmom} in the frictionless equation Eq.~\ref{eq:generalcyl} and Eq.~\ref{eq:nofric_eqs} corresponding to $i=1$, we have, after some simplifications:
\begin{eqnarray}
\frac{\ddot{\psi}}{\dot{\psi}} = -2 \frac{\dot{\phi}}{\phi} \nonumber \\
\dot{\psi} = \frac{b_2}{\phi^2}. \label{eq:nofric1}
\end{eqnarray}
where $b_2$ is a constant of integration, whose order is estimated in Appendix~\ref{app:b2scaling}. Substituting this into the $i=2$ frictionless equation and simplifying by neglecting all higher order $\phi$ terms, we eventually obtain 
\begin{equation}
\ddot{\phi} = \frac{A}{A+mR^2} \frac{b_2^2}{\phi^3}. \label{eq:nofric2}
\end{equation}
Using arguments identical to the pure-rolling case, this results in the following expression for the angle of turn, 
\begin{equation}
\Delta \psi_{turn} = \pi\sqrt{\frac{A+mR^2}{A}}. \label{eq:nofricangleofturn}
\end{equation}
for the frictionless limit.

\section{An alternate heuristic small angle treatment}
\label{sec:heuristic}
In this section, we derive the same angle of turn formulas (Eq.~\ref{eq:noslipangleofturn} and \ref{eq:nofricangleofturn}) without referring back to the complicated full dynamical equations. Rather, using heuristic reasoning, we directly derive equations of motion that apply at the small angle limit ($0 < \phi \ll 1$).

\par
We represent the lean of the cylinder by the vector $\Phi$ with magnitude equal to $\phi$ and direction along the $\mathbf{e}_{y1}$ axis: $\Phi = \phi \mathbf{e}_{y1}$. These quantities $\Phi, \phi$ and $\psi$ are related to each other exactly like $\mathbf{r}, r$ and $\theta$, respectively, in traditional polar coordinates. 

\par
Neglecting any angular velocity component along $\mathbf{e}_{z}$, the angular velocity vector for the cylinder is given by $\dot{\Phi} = \dot{\phi}\mathbf{e}_{y1} - \phi\dot{\psi}\mathbf{e}_{x1}$. The rate of change of angular momentum $\dot{\mathbf{H}}_G$ about the center of mass G is given by:
\begin{eqnarray}
\dot{\mathbf{H}}_G &=& A\ddot{\Phi} = A\frac{d^2}{dt^2}(\phi \mathbf{e}_{y1}) \nonumber \\
&=&  - A(\phi\ddot{\psi}+2\dot{\phi}\dot{\psi})\mathbf{e}_{x1} + A(\ddot{\phi}-\phi\dot{\psi}^2)\mathbf{e}_{y1}. \label{eq:heuristic_angmom}
\end{eqnarray}
The angular momentum balance equation is then given by 
\begin{equation}
\mathbf{\dot{H}}_G = \mathbf{M}_G, \label{eq:angmombalance}
\end{equation} 
where $\mathbf{M}_G$ is the moment of all the external forces about G. $\mathbf{M}_G$ depends on whether or not there is friction; so we treat the two cases in turn.

\subsection{Sliding without friction}
The vertical position of the center of mass is $z_G = R\sin(\phi) + H\cos(\phi)$. So $\ddot{z}_G \approx R \ddot{\phi}$. The vertical ground reaction is thus $m\ddot{z}_G = mR\ddot{\phi}$, neglecting gravity $mg$ in comparison. The moment $\mathbf{M}_G$ of this vertical force about the center of mass G is equal to $-m(R\cos{\phi}-H\sin{\phi})R\ddot{\phi} \approx -MR^2\ddot{\phi}$ in the direction $\mathbf{e}_{y1}$. Substituting this along with Eq.~\ref{eq:heuristic_angmom} in the angular momentum balance equation Eq.~\ref{eq:angmombalance}, we have
\begin{equation}
-MR^2\ddot{\phi}\mathbf{e}_{y1} = A(\ddot{\phi}-\phi\dot{\psi}^2)\mathbf{e}_{y1} - A(\phi\ddot{\psi}+2\dot{\phi}\dot{\psi})\mathbf{e}_{x1}. \nonumber
\end{equation}
This vector equation is identical to Eqs.~\ref{eq:nofric1} and~\ref{eq:nofric2} and therefore, lead to the same angle of turn (Eq.~\ref{eq:nofricangleofturn}).

\subsection{Rolling without slip}
The position of the center of mass G with respect to the point P$'$ on the cylinder in contact with ground is given by $\mathbf{r}_{G} \approx \mathbf{r}_{P'} - (R-H\phi)\mathbf{e}_{x1} + (R\phi + H)\mathbf{e}_{z}$. The velocity of the center of mass G is given by $\dot{\mathbf{r}}_{G} = \dot{\mathbf{r}}_{P'} + \dot{\Phi} \times (\mathbf{r}_G-\mathbf{r}_{P'})$. Using the no-slip constraint $\dot{\mathbf{r}}_{P'} = \mathbf{0}$, we obtain after some simplifications, the acceleration of the center of mass to first order to be 
\begin{equation}
\ddot{\mathbf{r}}_{G} = H(\ddot{\phi}-\phi\dot{\psi}^2)\mathbf{e}_{x1} + H(\phi\ddot{\psi}+2\dot{\phi}\dot{\psi})\mathbf{e}_{y1} + R\ddot{\phi}\mathbf{e}_{z}. \label{eq:rGdotdot}
\end{equation} 
The ground reaction force, which now includes a horizontal friction force as well, is simply $m\ddot{\mathbf{r}}_{G}$, neglecting gravity. The moment of this ground reaction force about G is given by 
\begin{eqnarray}
\mathbf{M}_G &=& (\mathbf{r}_{P'}-\mathbf{r}_G) \times m\ddot{\mathbf{r}}_{G} \nonumber \\
				&\approx& -\mathbf{e}_{x1} H^2(\phi\ddot{\psi} + 2\dot{\phi}\dot{\psi}) - \mathbf{e}_{y1}(R^2\ddot{\phi} + H^2(\ddot{\phi}-\phi\dot{\psi}^2))  \nonumber \\
					&\ &-\mathbf{e}_{z1}RH(\phi\ddot{\psi}+2\dot{\phi}\dot{\psi}). \nonumber
\end{eqnarray}

Equating $\mathbf{M}_G$ above to $\dot{\mathbf{H}}_{G}$  from Eq.~\ref{eq:heuristic_angmom} gives Eqs.~\ref{eq:noslip1} and ~\ref{eq:ddphianddpsi} from the previous version of the derivation, therefore resulting in the same formula for the angle of turn (Eq.~\ref{eq:noslipangleofturn}).

\section{Quantitative comparisons}
Firsly and most significantly, note that the limiting angle of turn does not depend on the initial conditions such as the initial tilt and tip velocity. This means that we do not have to control these accurately in an experiment. This also agrees with the relatively small variance in the histogram of Fig.~\ref{fig:scatterplotovaltine}, in which we did not control the initial conditions.

\par
The histogram Fig.~\ref{fig:scatterplotovaltine} was obtained using the cylindrical container shown in Fig.~\ref{fig:OvaltineTippedPhoto} with $R = 5.1$ cm, $H = 6.9$ cm, $A/m = 5.13 \times 10^{-3}$ m$^2$. Using these numbers in the angle of turn formulas gives an angle of turn of about 220 degrees for frictionless sliding and an angle of turn of $202$ degrees for rolling without slip. These angles of turn would manifest as a deviation of either $40$ degrees or $22$ degrees from falling over to exactly the diametrically opposite side. In the toppling experiment of Fig.~\ref{fig:scatterplotovaltine}, the container orientation was 33 degrees on average from 180 degrees in the leftward falls (suggesting an angle of turn of 213 degrees) and was about 37.9 degrees on average from 180 degrees in the rightward falls (suggesting angle of turn of 217.9 degrees). The standard deviations were respectively 3.9 and 4.5 degrees respectively. The experimental angle of turn seems better predicted by the asymptotic formula for frictionless sliding in this case. Although, neither the frictionless limit nor the no-slip limit is just right, both limits capture the many qualitative aspects of the motion quite well.

\section{Special limit: Tall thin containers}
For tall thin cylinders with $A \sim mH^2$ and $H >> R$, both equations for the angle of turn, Eq.~\ref{eq:noslipangleofturn} and Eq.~\ref{eq:nofricangleofturn}, tend to $\pi$ radians. That is, very tall cylinders are predicted to have a smaller symmetry-breaking. This prediction agrees with the common experience that when we tip a tall-enough cylinder (something that looks more like a tall thin beer bottle) in a manner that its bottom surface nearly falls flat, the cylinder essentially rocks up on a contact point almost diametrically opposite to the initial contact point. Thus, this apparently almost-symmetric rocking is accomplished by a rapid asymmetric rolling or sliding of the container over roughly one half of its bottom rim! This result is basically independent of the container-table frictional properties.

\section{Special limit: Disks} 
\label{sec:disks}
We may define a disk as a container with zero height: $H=0$, a good approximation is the Euler's disk \cite{Mof00}. If the radius of gyration of a disk is $R_g = kR$, then $C = mR^2k^2$ and $A = mR^2k^2/2$. Substituting these into the angle of turn formulas for the no-slip (Eq.~\ref{eq:noslipangleofturn}) or the frictionless (Eq.~\ref{eq:nofricangleofturn}) cases, we obtain the same angle of turn:
\begin{equation}
\Delta \psi_{turn} = \pi \sqrt{2+k^2}/{k}. \label{eq:diskangleofturn}
\end{equation}
Indeed, numerical exploration with Eqs.~\ref{eq:generalcyl}-\ref{eq:nofric_eqs}, suitably modified for frictional slip and specialized to disks, shows no dependence of the angle of turn $\Delta \psi_{turn} $ on the form of the friction law or the magnitude of the friction.  This lack of dependence on friction could be anticipated from our small angle calculations for pure rolling cylinders, specialized to disks. In particular, we find that the acceleration of the center of mass for a pure rolling cylinder  (Eq.~\ref{eq:rGdotdot}) with $H=0$ is vertical in the small-angle limit, enabling the no-slip condition to be satisfied even without friction. Thus, the rolling solution is obtained with or without friction. For the special case of pure-rolling disks, Eq.~\ref{eq:diskangleofturn} was found in  \cite{Cus06}.

\par
\textbf{A homogeneous disk} has $k = 1/\sqrt{2}$. The corresponding angle of turn is equal to $\pi\sqrt{5} \approx 2.23 \pi$, which is about 41 degrees more than a full rotation of the contact point. This prediction is easily confirmed in casual experimentation with metal caps of large-mouthed bottles or jars on sturdy tables -- for such caps, we observe that the new contact point is invariably quite close to the old contact point.

\par
\textbf{A ring} such as the rim of a bicycle wheel has $k \approx 1$. The corresponding angle of turn is equal to $\pi\sqrt{3} \approx 1.73 \pi$, which is about 48 degrees less than a full rotation of the contact point. Thus the apparent near-collisional behavior of a homogeneous disk and a ring will be superficially similar, even though the actual angles of turn differ by about $90$ degrees. 

\par
The angle of turn can be controlled by adjusting $k$. For instance, between a ring and a disk is an object that appears to bounce straight back up the way it falls. And it is possible to increase the theoretical angle of turn without bound by choosing $k \rightarrow 0$, a disk in which almost all the mass is concentrated at the center.

\section{Summary}
Here we analyzed what happens when a cylinder or a disk rocks to an almost flat collision on its bottom surface. We found that the smallest deviation from a perfect face-down collision of a container's bottom results in a rapid rolling and/or sliding motion in which the contact point moves through a finite angle in infinitesimal time. Calculations of this finite angle explain certain apparent symmetry breaking in experiments involving rocking or toppled containers.

\par
In this system,  the  consequences of such a degenerate `collision' are a discontinuous
dependence on initial conditions (rolling left or rolling right depend on the smallest
deviations in the initial conditions). Such discontinuous dependence on initial
conditions or geometry is a generic feature of systems in the neighborhood
of simultaneous collisions. Other examples include a pool break or the 
rolling polygon of \cite{Rui05}.

\begin{acknowledgments}
Thanks to Arend Schwab for comments on an early manuscript and Mont Hubbard for editorial comments.
The alternate heuristic derivation of Section~\ref{sec:heuristic} were informed by
discussions with Anindya Chatterjee in the context of the Euler's disk. 
MS was supported by Cornell University in the early stages of this work (2000).
The work was also supported partly by NSF robotics (CISE-0413139) and FIBR (EF-0425878) grants.
\end{acknowledgments}

\appendix

\section{Deriving equations of motion}
\label{app:eqmotion}
The moment of inertia tensor of the cylinder about its center of mass is, in dyadic form, 
\begin{equation}
\mathbf{I}_{G} = A (\mathbf{e}_{x2} \otimes \mathbf{e}_{x2}) + A (\mathbf{e}_{y2} \otimes \mathbf{e}_{y2}) + C (\mathbf{e}_{z2} \otimes \mathbf{e}_{z2}).
 \end{equation}
 The angular velocity $\mathbf{\Omega}$ of the cylinder is given by 
\begin{equation}
\mathbf{\Omega} = \dot{\psi}\sin{\phi}\,\mathbf{e}_{x2} - \dot{\phi}\,\mathbf{e}_{y2} + (\dot{\theta}+\dot{\psi}\cos{\phi})\,\mathbf{e}_{z2}.
\end{equation}
The angular momentum about the center of mass is 
\begin{eqnarray}
\mathbf{H}_{/G} &=& \mathbf{I}_G\ \mathbf{\Omega} \nonumber \\
	&=&  A\dot{\psi}\sin{\phi} \,\mathbf{e}_{x2} - A\dot{\phi}\, \mathbf{e}_{y2} + C(\dot{\theta}+\dot{\psi}\cos{\phi}) \,\mathbf{e}_{z2} \nonumber \\
	\label{eq:HG}
\end{eqnarray}
The rate of change of angular momentum $\mathbf{H}_{/G}$ is given by the sum of two terms: 1) the rate of change relative to the rotating frame $\mathbf{e}_{x2}$-$\mathbf{e}_{y2}$-$\mathbf{e}_{z2}$, obtained by simply differentiating the components in Eq.~\ref{eq:HG}, and 2) the rate of change $\mathbf{\Omega}_2 \times \mathbf{H}_{/G}$ due to the rotation of the $\mathbf{e}_{x2}$-$\mathbf{e}_{y2}$-$\mathbf{e}_{z2}$ frame with an angular velocity $\mathbf{\Omega}_2 = \dot{\psi}\sin{\phi}\,\mathbf{e}_{x2} - \dot{\phi}\,\mathbf{e}_{y2} + \dot{\psi}\cos{\phi}\,\mathbf{e}_{z2}$.
\begin{eqnarray}
&\mathbf{\dot{H}}_{/G} = \left(2A\dot{\psi}\dot{\phi}\cos{\phi}+A\ddot{\psi}\sin{\phi}-C\dot{\phi}(\dot{\theta}+\dot{\psi}\cos{\phi})  \right) \mathbf{e}_{x2} \nonumber \\
&+ \left(-A\ddot{\phi}+A\dot{\psi}^2\sin{\phi}\cos{\phi}-C\dot{\psi}\sin{\phi}(\dot{\theta}+\dot{\psi}\cos{\phi}) \right) \mathbf{e}_{y2} \nonumber \\
&+ \ C\left(-\dot{\psi}\dot{\phi}\sin{\phi}+\ddot{\psi}\cos{\phi}+\ddot{\theta} \right) \mathbf{e}_{z2}
\end{eqnarray}

\subsection{No-slip rolling}
The point S on the cylinder in contact with the ground has zero velocity:  $\mathbf{v}_S = \mathbf{0}$, no slip. 
Using $\mathbf{r}_{SG} = \mathbf{r}_{PG} = R\mathbf{e}_{x2}+ H \mathbf{e}_{z2}$, the velocity 
$\mathbf{v}_G$ of the center of mass is given by
\begin{eqnarray}
\mathbf{v}_G &=& \mathbf{v}_S + \mathbf{\Omega} \times \mathbf{r}_{PG} \nonumber \\
&=& \mathbf{0}-H\dot{\phi} \mathbf{e}_{x2} + R \dot{\phi} \mathbf{e}_{z2}\nonumber \\
 &\ & + \left( R\dot{\theta}+R\dot{\psi}\cos{\phi}-H\dot{\psi}\sin{\phi} \right) \mathbf{e}_{y2}. \label{eq:vG_noslip}
\end{eqnarray}

The acceleration of the center of mass $\mathbf{a}_G$ is given by adding two terms: 1) acceleration relative to the rotating $\mathbf{e}_{x2}$-$\mathbf{e}_{y2}$-$\mathbf{e}_{z2}$ frame, obtained by differentiating the components in Eq.~\ref{eq:vG_noslip}, and 2) an acceleration term $\mathbf{\Omega} \times \mathbf{v}_G$ due to the rotation of $\mathbf{e}_{x2}$-$\mathbf{e}_{y2}$-$\mathbf{e}_{z2}$ frame. We obtain
\begin{eqnarray}
 \mathbf{a}_G &= &  d\mathbf{v}_G/dt \nonumber \\ 
 &=& (-H\ddot{\phi}-R\dot{\phi}^2-R\dot{\psi}\cos{\phi}\dot{\theta}-R\dot{\psi}^2\cos^2{\phi} \nonumber \\
&\ & +H\dot{\psi}^2\cos{\phi}\sin{\phi} ) \mathbf{e}_{x2} \nonumber \\
 &\ & + ( -2R\dot{\psi}\dot{\phi}\sin{\phi}-2H\dot{\psi}\dot{\phi}\cos{\phi}+R\ddot{\psi}\cos{\phi} \nonumber \\
 &\ &-H\ddot{\psi}\sin{\phi}+R\ddot{\theta} ) \mathbf{e}_{y2} \nonumber \\
 &\ &+ ( R\ddot{\phi}+R\dot{\psi}\dot{\theta}\sin{\phi}+R\dot{\psi}^2\sin{\phi}\cos{\phi} \nonumber \\
 &\ & -H\dot{\psi}^2\sin^2{\phi}-H\dot{\phi}^2 ) \mathbf{e}_{z2}.
\end{eqnarray}

The contact force $\mathbf{F}_P$ is given by linear momentum balance:
\begin{equation}
\mathbf{F}_P = m\mathbf{a}_G + mg \mathbf{e}_{z}.
\end{equation} 
Then, the moment $\mathbf{M}_{/G}$ of all the external forces about the center of mass is 
\begin{eqnarray}
\mathbf{M}_{/G} &=& \mathbf{r}_{GP} \times \mathbf{F}_P = -\mathbf{r}_{PG} \times \mathbf{F}_P \nonumber \\
				&=& B_1 \mathbf{e}_{x2} + B_2 \mathbf{e}_{y2} + B_3 \mathbf{e}_{z2},
\end{eqnarray}				                                                                            
in which, upon simplification, we have
\begin{eqnarray}
B_1 &=& mH(-2R\dot{\psi}\dot{\phi}\sin{\phi}-2H\dot{\psi}\dot{\phi}\cos{\phi} \nonumber \\
&+& R\ddot{\psi}\cos{\phi}-H\ddot{\psi}\sin{\phi}+R\ddot{\theta}) \nonumber \\
B_2 &=& m(-gH\sin{\phi}+H^2\ddot{\phi}+HR\dot{\psi}\dot{\theta}\cos{\phi}+2HR\dot{\psi}^2\cos^2{\phi} \nonumber \\
&-&H^2\dot{\psi}^2\cos{\phi}\sin{\phi}+gR\cos{\phi}+R^2\ddot{\phi}+R^2\dot{\psi}\dot{\theta}\sin{\phi} \nonumber \\
&+&R^2\dot{\psi}^2\sin{\phi}\cos{\phi}-RH\dot{\psi}^2) \nonumber \\
B_3 &=& -mR(-2R\dot{\psi}\dot{\phi}\sin{\phi}-2H\dot{\psi}\dot{\phi}\cos{\phi} \nonumber \\
&+& R\ddot{\psi}\cos{\phi}-H\ddot{\psi}\sin{\phi}+R\ddot{\theta}). \nonumber 
\end{eqnarray}

Equating $\mathbf{M}_{/G}$ and $\dot{\mathbf{H}}_{/G}$ gives us the angular momentum balance equations of motion for no-slip rolling in Eqs.~\ref{eq:generalcyl},\ref{eq:noslip_eqs}.

\subsection{Frictionless sliding}
No friction implies that the horizontal velocity of the center of mass G is a constant and can be set to zero by appropriate reference frame choice, without loss of generality. Further, noting that $z_G = R\sin{\phi} +H\cos{\phi}$, we have, by differentiating twice:
\begin{align}
&\mathbf{a}_G = \ddot{z}_G \ \mathbf{e}_z \nonumber \\
&= \left( (R\cos{\phi}-H\sin{\phi})\ddot{\phi} -(R\sin{\phi}+H\cos{\phi})\dot{\phi}^2 \right)  \mathbf{e}_z \nonumber \\
&= \sin{\phi}\left((R\cos{\phi}-H\sin{\phi})\ddot{\phi}-(R\sin{\phi}+H\cos{\phi})\dot{\phi}^2\right) \mathbf{e}_{x2} \nonumber \\
&+\cos{\phi}\left((R\cos{\phi}-H\sin{\phi})\ddot{\phi}-(R\sin{\phi}+H\cos{\phi})\dot{\phi}^2\right)  \mathbf{e}_{z2} \nonumber 
\end{align}

As before, we compute the contact force as $\mathbf{F}_P = m\mathbf{a}_G + mg\mathbf{e}_{z}$ and compute the net moment $\mathbf{M}_{G}$ about the center of mass as 
\begin{eqnarray}
\mathbf{M}_{/G} &=& \mathbf{r}_{GP} \times \mathbf{F}_P \nonumber \\
&=& m(R\cos{\phi}-H\sin{\phi})(-R\dot{\phi}^2\sin{\phi} \nonumber \\
&+&R\ddot{\phi}\cos{\phi}-H\dot{\phi}^2\cos{\phi}-H\ddot{\phi}\sin{\phi}+g) \mathbf{e}_{y2} \nonumber 
\end{eqnarray}
Again, equating $\mathbf{M}_{/G}$ with the $\dot{\mathbf{H}}_{/G}$ before gives us the required equations of motion Eqs.~\ref{eq:generalcyl},\ref{eq:nofric_eqs}.

\section{Scaling of $b_2$}
\label{app:b2scaling}
\subsection{No-slip rolling}
We now establish that $b_2 = O(\phi_c^2)$, used in obtaining Eq.~\ref{eq:noslipangleofturn}. The total mechanical energy $E$ of the cylinder, a constant, is given by:
\begin{eqnarray}
2E &=& m(\mathbf{v}_G \cdot \mathbf{v}_G) + \mathbf{\Omega} \cdot \mathbf{I}_G\mathbf{\Omega} + \textrm{P. E.} \\
	&=& m\dot{\phi}^2(H^2+R^2) + C(\dot{\theta}+\dot{\psi}\cos{\phi})^2 \nonumber \\
  &\ & + m(R\dot{\theta}+R\dot{\psi}\cos{\phi} - H\dot{\psi}\sin{\phi})^2  \nonumber \\
  &\ & + A\dot{\phi}^2 + A\dot{\psi}^2\sin^2{\phi} + \textrm{P.E.} 
\end{eqnarray}
in which P.E. is the potential energy. We consider this total energy at the time of lowest tip angle $\phi$, $t = t_c$, we have $\dot{\phi} = 0$, $\phi = \phi_c$, $\dot{\psi} = b_2/\phi_c^2$, and the potential energy P.E. $ = mg(R\sin{\phi}+H\cos{\phi}) = \textrm{Constant}+O(\phi_c)$. Making use of the usual small $\phi$ approximations (as used in the main text) and Eq.~\ref{eq:psieqtheta}, namely $\dot{\psi}+\dot{\theta} \sim \dot{\psi}\phi$, we find that 
\begin{equation}
E \sim \dot{\psi}^2 \phi_c^2 = \left(\frac{b_2}{\phi_c^2} \right)^2\phi_c^2. \\
\end{equation}
Because $E$ is a constant, we have $b_2^2 = O(\phi_c^2)$ or $b_2 = O(\phi_c)$ as claimed earlier.

\subsection{Frictionless sliding}
The total mechanical energy $E$ is now given by 
\begin{eqnarray}
2E &=&  m\dot{\phi}^2(R\cos{\phi}-H\sin{\phi})^2 \nonumber \\
&\ &+ A \dot{\psi}^2\sin^2{\phi} + A \dot{\phi}^2 + C(\dot{\theta} +\psi\cos{\phi})^2 \nonumber \\
&\ &  + mg(R\sin{\phi}+H\cos{\phi}).
\end{eqnarray}
Again, using small $\phi_c$ approximations and considering the energy equation at $t=t_c$, we obtain $b_2 = O(\phi_c)$, using arguments identical to the no-slip rolling case.


\begin{thebibliography}{7}
\expandafter\ifx\csname natexlab\endcsname\relax\def\natexlab#1{#1}\fi
\expandafter\ifx\csname bibnamefont\endcsname\relax
  \def\bibnamefont#1{#1}\fi
\expandafter\ifx\csname bibfnamefont\endcsname\relax
  \def\bibfnamefont#1{#1}\fi
\expandafter\ifx\csname citenamefont\endcsname\relax
  \def\citenamefont#1{#1}\fi
\expandafter\ifx\csname url\endcsname\relax
  \def\url#1{\texttt{#1}}\fi
\expandafter\ifx\csname urlprefix\endcsname\relax\def\urlprefix{URL }\fi
\providecommand{\bibinfo}[2]{#2}
\providecommand{\eprint}[2][]{\url{#2}}

\bibitem[{\citenamefont{Ivanov}(1995)}]{Ivan95}
\bibinfo{author}{\bibfnamefont{A.~P.} \bibnamefont{Ivanov}},
  \bibinfo{journal}{Journal of Applied Mathematics and Mechanics}
  \textbf{\bibinfo{volume}{59}}, \bibinfo{pages}{887} (\bibinfo{year}{1995}).

\bibitem[{\citenamefont{Goyal et~al.}(1998)\citenamefont{Goyal, Papadopoulos,
  and Sullivan}}]{Goy98}
\bibinfo{author}{\bibfnamefont{S.}~\bibnamefont{Goyal}},
  \bibinfo{author}{\bibfnamefont{J.~M.} \bibnamefont{Papadopoulos}},
  \bibnamefont{and} \bibinfo{author}{\bibfnamefont{P.~A.}
  \bibnamefont{Sullivan}}, \bibinfo{journal}{Journal of Dynamic Systems,
  Measurement, and Control} \textbf{\bibinfo{volume}{120}}, \bibinfo{pages}{83}
  (\bibinfo{year}{1998}).

\bibitem[{\citenamefont{Chatterjee and Ruina}(1998)}]{Chat98}
\bibinfo{author}{\bibfnamefont{A.}~\bibnamefont{Chatterjee}} \bibnamefont{and}
  \bibinfo{author}{\bibfnamefont{A.}~\bibnamefont{Ruina}},
  \bibinfo{journal}{ASME Journal of Applied Mechanics}
  \textbf{\bibinfo{volume}{65}}, \bibinfo{pages}{939} (\bibinfo{year}{1998}).

\bibitem[{\citenamefont{Ruina et~al.}(2005)\citenamefont{Ruina, Bertram, and
  Srinivasan}}]{Rui05}
\bibinfo{author}{\bibfnamefont{A.}~\bibnamefont{Ruina}},
  \bibinfo{author}{\bibfnamefont{J.~E.~A.} \bibnamefont{Bertram}},
  \bibnamefont{and}
  \bibinfo{author}{\bibfnamefont{M.}~\bibnamefont{Srinivasan}},
  \bibinfo{journal}{Journal of Theoretical Biology}
  \textbf{\bibinfo{volume}{237}}, \bibinfo{pages}{170} (\bibinfo{year}{2005}).

\bibitem[{\citenamefont{O'Reilly}(1996)}]{Orei96}
\bibinfo{author}{\bibfnamefont{O.~M.} \bibnamefont{O'Reilly}},
  \bibinfo{journal}{Nonlinear Dynamics} \textbf{\bibinfo{volume}{10}},
  \bibinfo{pages}{287} (\bibinfo{year}{1996}).

\bibitem[{\citenamefont{Borisov and Mamaev}(2002)}]{Bor03}
\bibinfo{author}{\bibfnamefont{A.~V.} \bibnamefont{Borisov}} \bibnamefont{and}
  \bibinfo{author}{\bibfnamefont{I.~S.} \bibnamefont{Mamaev}},
  \bibinfo{journal}{Regular and Chaotic Dynamics} \textbf{\bibinfo{volume}{7}},
  \bibinfo{pages}{177} (\bibinfo{year}{2002}).

\bibitem[{\citenamefont{Cushman and Duistermaat}(2006)}]{Cus06}
\bibinfo{author}{\bibfnamefont{R.~H.} \bibnamefont{Cushman}} \bibnamefont{and}
  \bibinfo{author}{\bibfnamefont{J.~J.} \bibnamefont{Duistermaat}},
  \bibinfo{journal}{Regular and Chaotic Dynamics}
  \textbf{\bibinfo{volume}{11}}, \bibinfo{pages}{31} (\bibinfo{year}{2006}).
  
 \bibitem[{\citenamefont{Moffatt}(2000)}]{Mof00}
\bibinfo{author}{\bibfnamefont{H.~K.} \bibnamefont{Moffatt}},
  \bibinfo{journal}{Nature}
  \textbf{\bibinfo{volume}{404}}, \bibinfo{pages}{833} (\bibinfo{year}{2000}). 
  
\end{thebibliography}
\end{document}